\documentclass[%
 reprint,
superscriptaddress,
 amsmath,amssymb,
 aps,
 prl,
floatfix,
]{revtex4-2}
\usepackage{csquotes}
\usepackage[utf8]{inputenc}
\usepackage{graphicx}
\usepackage{dcolumn}
\usepackage{bm}
\usepackage{xcolor}
\usepackage{microtype}
\usepackage[
    separate-uncertainty = true,
    ]{siunitx}
\usepackage{cleveref}

\begin{document}

\title{All optical control of bubble and skyrmion breathing}

\author{Tim Titze}
\affiliation{I.\,Physikalisches Institut, Universit\"at G\"ottingen, 37077 G\"ottingen, Germany\looseness=-1}
\author{Timo Schmidt}
\affiliation{Institute of Physics, University of Augsburg, 86135 Augsburg, Germany\looseness=-1}
\author{Manfred Albrecht}
\affiliation{Institute of Physics, University of Augsburg, 86135 Augsburg, Germany\looseness=-1}
\author{Stefan Mathias}
\email[]{smathias@uni-goettingen.de}
\affiliation{I.\,Physikalisches Institut, Universit\"at G\"ottingen, 37077 G\"ottingen, Germany\looseness=-1}
\affiliation{International Center for Advanced Studies of Energy Conversion (ICASEC), Universit\"at G\"ottingen, 
37077 G\"ottingen, Germany} 
\author{Daniel Steil}
\email[]{dsteil@gwdg.de}
\affiliation{I.\,Physikalisches Institut, Universit\"at G\"ottingen, 37077 G\"ottingen, Germany\looseness=-1}

\begin{abstract}
    Controlling the dynamics of topologically protected spin objects by all optical means promises enormous potential for future spintronic applications. Excitation of bubbles and skyrmions in ferrimagnetic [Fe(0.35 nm)/Gd(0.40 nm)]$_{160}$ multilayers by ultrashort laser pulses leads to a periodic modulation of the core diameter of these spin objects, the so-called breathing mode. We demonstrate versatile amplitude and phase control of this breathing using a double excitation scheme, where the observed dynamics is controlled by the excitation delay. We gain insight into both the time scale on which the breathing mode is launched and the role of the spin object size on the dynamics. Our results demonstrate that ultrafast optical excitation allows for precise tuning of the spin dynamics of trivial and non-trivial spin objects, showing a possible control strategy in device applications.  
\end{abstract}

\maketitle

Topologically protected magnetic spin textures are expected to be a key building block for future applications in spintronics and unconventional computing techniques such as neuromorphic computing~\cite{Finocchio2016, Fert2017, Prychynenko2018, Song2020, Zhang2020, Chumak2022}. Magnetic skyrmions, i.e., magnetic whirls, characterized by their intricate spin configuration and topologically non-trivial nature~\cite{Bogdanov1994, Roessler2006, Yu2010}, are in the center of current research efforts to translate fundamental science into future devices~\cite{Lenk2011, Yu2021, Petti2022}. Therefore, magnetic, electrical and microwave manipulation of skyrmions has been the subject of intense research in recent years~\cite{Finocchio2016, Fert2017, Lonsky2020, Wang2022}. Moreover, utilizing ultrashort laser pulses, several works uncovered the possibility of optical detection of magnetic skyrmions~\cite{Ogawa2015, Padmanabhan2019, Sekiguchi2022, Kalin2022,Titze2024} and even optical creation of magnetic skyrmions from various types of spin textures~\cite{Eggebrecht2017, Je2018, Buettner2021, Khela2023, Titze2024}.

In our work, we go a step beyond the optical detection and creation of localized spin objects and demonstrate in a two-pulse experiment that we are able to control the so-called breathing mode of bubbles and skyrmions, i.e., a periodic expansion and shrinking of localized spin objects in amplitude and phase depending on temporal delay between the two excitation pulses. Such an approach was indeed already proposed by Wang et al.~\cite{Wang2023} using a microwave driving field to achieve a coherent stimulated amplification of the skyrmion breathing mode, however, in our case, we achieve such control with all-optical pulses.

\begin{figure}[tb!]
     \centering
     \includegraphics[width=\columnwidth]{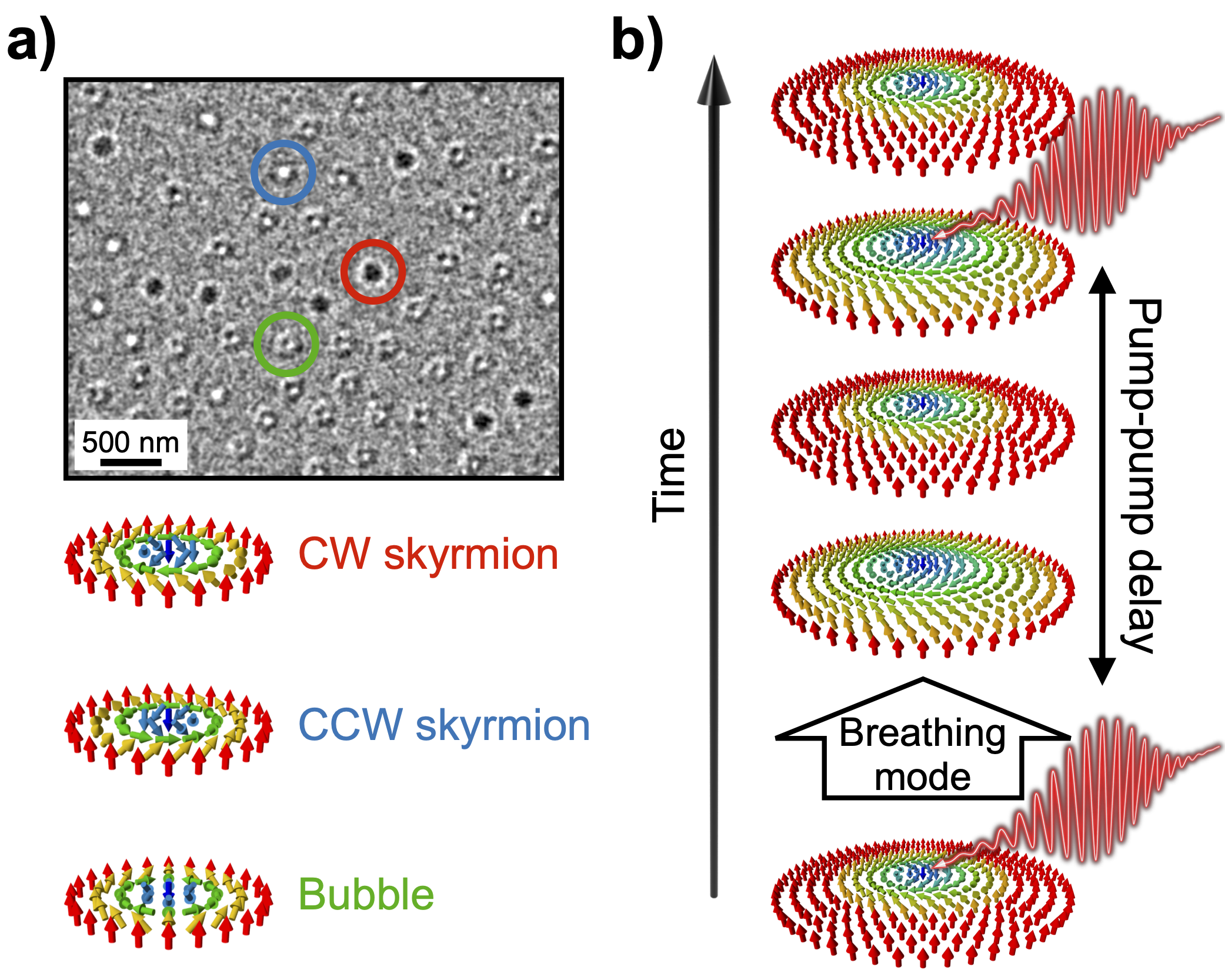}
     \caption{\textbf{a)} Lorentz transmission electron microscopy image of the bubble and skyrmion lattice in the Fe/Gd multilayer system. Highlighted are a clockwise (CW) and counterclockwise (CCW) Bloch-type skyrmion and one magnetic bubble with their respective spin structure given below. \textbf{b)} Optical excitation of a skyrmion induces the skyrmion breathing mode, i.e., a periodic modulation of its core diameter. A second optical excitation with variable time delay is used to modify the B/SK breathing mode oscillation at variable time delays or phase states.}
     \label{fig:Intro}
\end{figure}

We study a [Fe(0.35 nm)/Gd(0.40 nm)]$_{160}$ multilayer system containing a dense bubble and skyrmion (B/SK) lattice (see Fig.~\ref{fig:Intro}a) stabilized by dipolar interactions at moderate magnetic fields of $\mu_0 H=190-240$\,mT and at ambient temperature. For further sample information we refer to~\cite{Titze2024}. In~\cite{Titze2024}, we have shown that we can identify different magnetic spin textures in this material using the time-resolved magneto-optical Kerr effect (TR-MOKE), where the breathing mode of B/SKs manifests itself as a specific time-dependent oscillation of the out-of-plane magnetization. In the present study, we exploit these findings and aim to coherently control optically induced B/SK breathing dynamics using a double-pump excitation scheme as exemplarily shown in Fig.~\ref{fig:Intro}b for a single Bloch-type skyrmion. 

Here, the first optical pump excitation is used to start the collective B/SK breathing with a mode frequency of $f_{bsk}\approx 1.4$\,GHz for our material system. The second pump excitation is delayed and modifies the already started B/SK breathing mode oscillation at different phases of the B/SK breathing. The total spin dynamics induced by the double pump excitation is then probed using TR-MOKE. Briefly, magnetization dynamics $\Delta M(t)$ were measured using a bichromatic pump-probe setup using 1030\,nm pump pulses and 515\,nm probe pulses of less than 40\,fs pulse duration at a pulse repetition rate of 50\,kHz (for details, see~\cite{Titze2024}). In addition, the pump pulse was split into two, separately controlled by delay stages, to vary the pump-probe and the pump-pump delay independently.  

\begin{figure}[tb]
     \centering
     \includegraphics[width=\columnwidth]{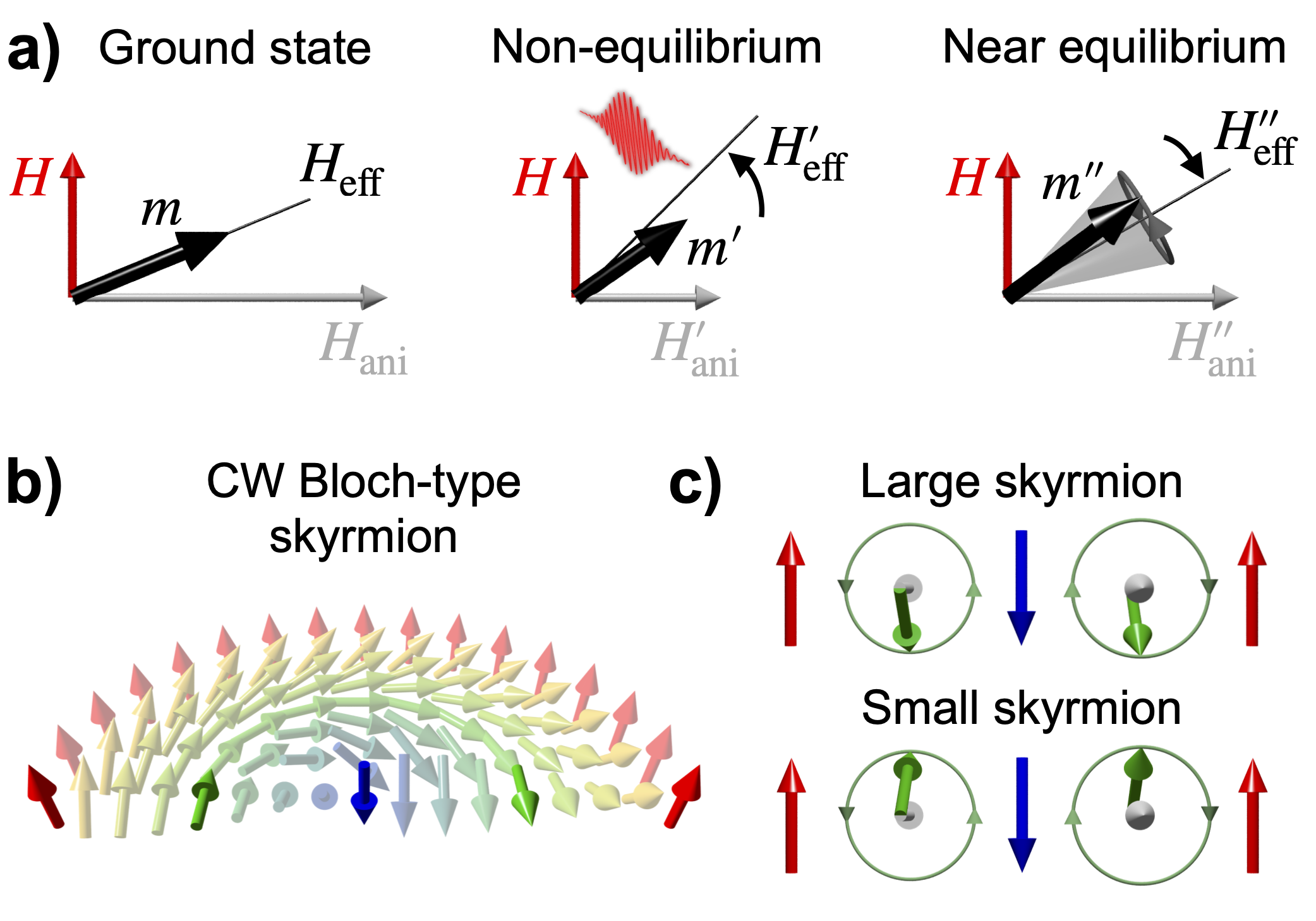}
     \caption{\textbf{a)} Optical excitation of magnetization precession. In the ground state $\boldsymbol{m}$ and $\boldsymbol{H_{\mathrm{eff}}}$ are aligned parallel. Laser-excitation into an initial strong non-equilibrium changes $\boldsymbol{m}$ and $\boldsymbol{H_{\mathrm{eff}}}$ to $\boldsymbol{m'}$ and $\boldsymbol{H'_{\mathrm{eff}}=H+H'_\textrm{ani}}$. On the few ps timescale a mostly recovered $\boldsymbol{m''\approx|m|}$ precesses around  $\boldsymbol{H''_{\mathrm{eff}}}$, which recovers on ns timescales to  $\boldsymbol{H_{\mathrm{eff}}}$ by thermal diffusion. \textbf{b)} Schematic depiction of half a clockwise Bloch-type skyrmion. \textbf{c)} Few spin representation of cross section highlighted in b) showcasing how precession leads to periodic breathing of the skyrmion core. Here, a large skyrmion core is shown at the top and a small skyrmion core at the bottom, as indicated by the precessing spins in green with their equilibrium direction in gray. The external field $\boldsymbol{H}$ points in the direction of the red arrows.}
     \label{fig:ExcitationModel}
\end{figure}

The collective B/SK breathing can be understood in terms of spin precession, showcased by the toy model in Fig.~\ref{fig:ExcitationModel}.
In general (see Fig.~\ref{fig:ExcitationModel}a), precession may occur when an ultrafast laser excitation heats up a ferromagnetic material, quenching the magnetization $\boldsymbol{m}$ on a sub-picosecond timescale~\cite{Beaurepaire1996}. Since the magnetic anisotropy strongly depends on temperature and magnetization, such a laser pulse leads to an ultrafast change of the effective anisotropy field $\boldsymbol{H_{\mathrm{eff}}}$~\cite{vanKampen2002,Kirilyuk2010} (consisting, e.g., of the external field $\boldsymbol{H}$, the demagnetizing field, and the magneto-crystalline anisotropy field), altering $\boldsymbol{H_{\mathrm{eff}}}$ to $\boldsymbol{H'_{\mathrm{eff}}}$. In case that the quenched $\boldsymbol{m'}$ and $\boldsymbol{H'_{\mathrm{eff}}}$ are not anymore parallel after the excitation, $\boldsymbol{m'}$ reacts by reorienting during the initial strong nonequilibrium~\cite{Vomir2005,Bigot2005, Vomir2006}. Afterwards, $\boldsymbol{m'}$ and $\boldsymbol{H'_{\mathrm{eff}}}$ typically recover within a few picoseconds close to their original values, but different directions of both lead to a torque on $\boldsymbol{m''}$ inducing precessional motion on longer timescales~\cite{vanKampen2002, Bigot2005, Kirilyuk2010, Rzhevsky2007}.
With respect to a skyrmion as shown in Fig.~\ref{fig:ExcitationModel}b), such precessional motion translates into an expansion and shrinking of the skyrmion core, see the simplified cross sections~\cite{Ehlers2016} in Fig.~\ref{fig:ExcitationModel}c). In the present experiment this precession is induced by (temperature-driven) changes of sample-intrinsic anisotropy terms $\boldsymbol{H_\textrm{ani}}$ due to laser excitation, whereas the external field term stays constant leading to a time-dependent tilt of $\boldsymbol{H_{\mathrm{eff}}}$ compared to the equilibrium case.  

\begin{figure}[t]
     \centering
     \includegraphics[width=\columnwidth]{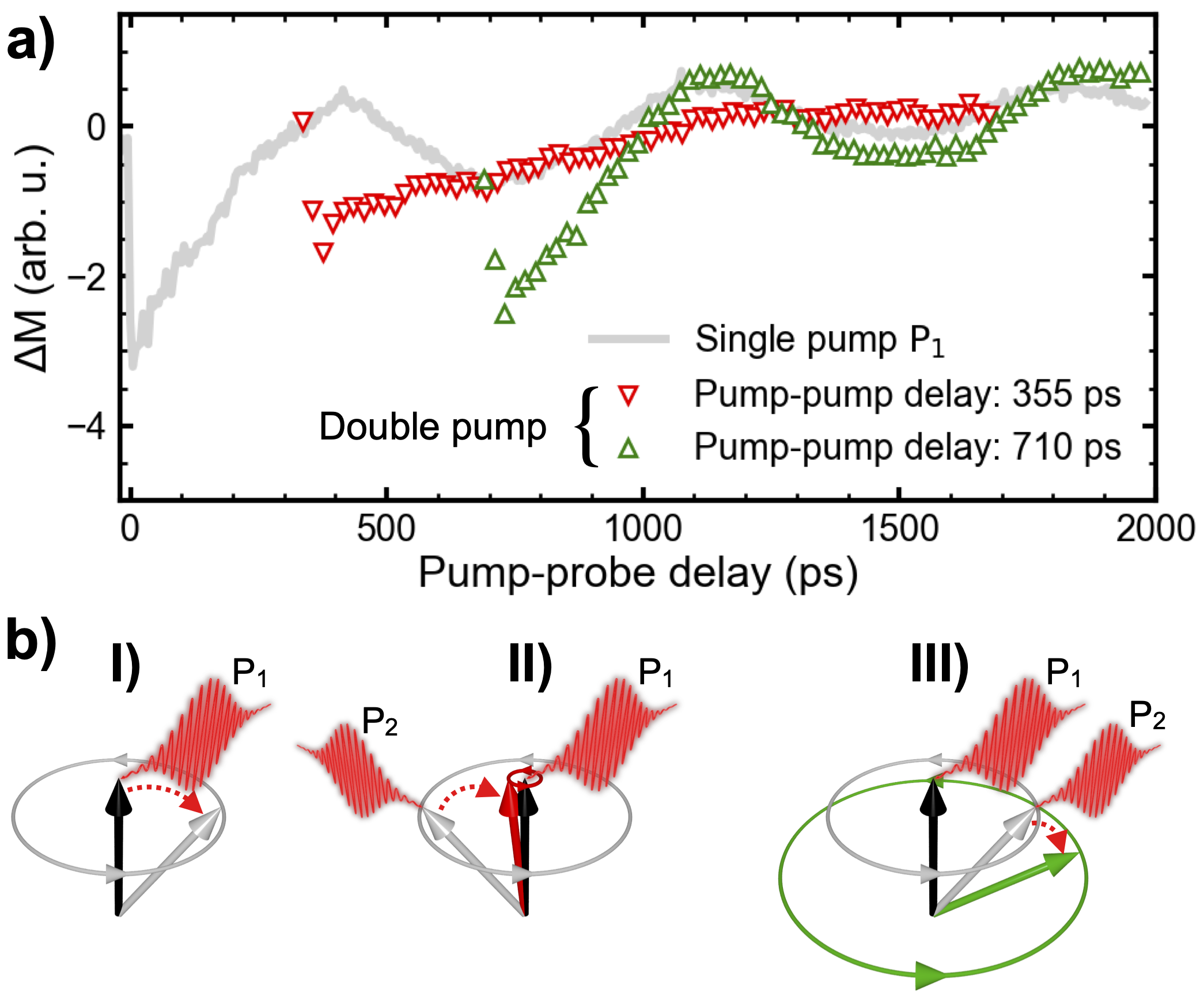}
     \caption{\textbf{a)} Magnetization dynamics in a double pump excitation scheme for different pump-pump delays of 355\,ps (red) and 710\,ps (green), respectively. Here, the pump-probe delay is referenced to the arrival of the first pump $P_1$. \textbf{b)} Phasor representation of an arbitrary, non-oop spin upon laser excitation. Single excitation I) causes magnetization precession (gray arrow). Double excitation either decreases II) (red arrow) or increases III) (green arrow) the precession amplitude, depending on the pump-pump delay.}
     \label{fig:DoublePumpSignal}
\end{figure}

Figure~\ref{fig:DoublePumpSignal}a) depicts the measured magnetization dynamics resulting from a double pump excitation of a B/SK lattice stabilized by a magnetic field of $\mu_0 H=193$\,mT in out-of-plane (oop) direction for two different pump-pump delays of 355\,ps (red) and 710\,ps (green). These delays correspond to a second excitation at $\phi=\pi$ and $\phi=2\pi$ precession period of the B/SK breathing mode, respectively. Furthermore, the response to only the first excitation ($P_1$) is plotted as a gray line~\footnote{The response to a single pump pulse, either $P_1$ or $P_2$, can be directly extracted from the double pump measurement, if either $P_2$ or $P_1$  arrive after the probe pulse, i.e., for about 20\,$\mu$s pump-pump delay. In this case, the energy of the pump pulse arriving after the probe pulse is already dissipated from the system prior to the arrival of the pump pulse initiating the dynamics measured by the probe.}. Note that the fluence of the second pulse ($F\approx 0.4\,\mathrm{mJ/cm^2}$) is chosen to be slightly less than that of the first pulse ($F\approx 0.7\,\mathrm{mJ/cm^2}$) to adjust for the time-dependent damping of the amplitude of the breathing mode.

Clearly, we find that the overall dynamic response shows a strong dependence on the phase delay of the double-pump excitation. At 355\,ps after the first excitation, corresponding to a phase of $\pi$ of the breathing mode, the system reaches a state of maximum magnetization, which means that the diameter of the B/SK spin objects is minimized. In contrast, the spin objects are in a state of maximum core size at 710\,ps corresponding to a breathing mode phase of $2\pi$, which results in a minimum in the total magnetization. Further excitation at a phase delay of $2\pi$ or $\pi$  then leads to an amplification or almost full attenuation of the B/SK breathing mode depending on whether the spin objects have had maximum  (710\,ps) or minimum (355\,ps) size at the time of the second pump excitation, respectively. We explain our findings using a phasor representation of the magnetization depicted in Figure~\ref{fig:DoublePumpSignal}b) similar to~\cite{Berk2019}. As already mentioned (see Fig.~\ref{fig:ExcitationModel}), laser excitation triggers a spin precession I) indicated by the gray arrow in Fig.~\ref{fig:DoublePumpSignal}b) by changing $\boldsymbol{H_\textrm{ani}}$. A second pump pulse can then either suppress II) or enhance III) the precession amplitude by a similar anisotropy change depending on the pump-pump delay or precession phase.

\begin{figure}[tb]
     \centering
     \includegraphics[width=\columnwidth]{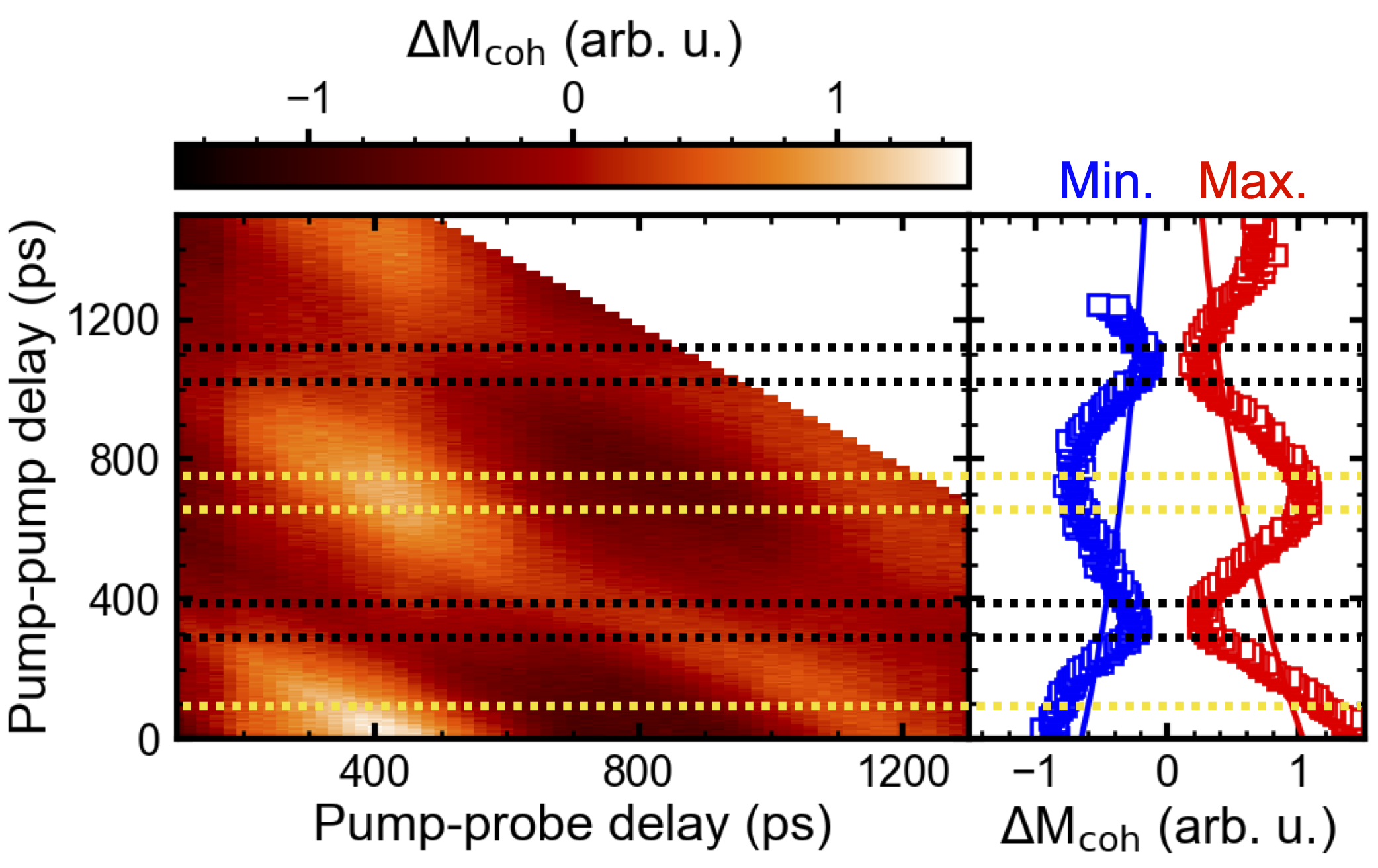}
     \caption{\textbf{Left:} Coherent magnetization dynamics in dependence of pump-pump and pump-probe delay referenced to the second pump $P_2$. \textbf{Right:} Both the maximum (red squares) and minimum (blue squares) signal for the full range of pump-probe delays is plotted against the pump-pump delay in comparison to the single excitation (red and blue lines) taking damping into account. Regions of pump-pump delays at which either maximum amplification (yellow) or attenuation of the breathing mode occurs (black) are highlighted.}
     \label{fig:2DPlot}
\end{figure}

To analyze the breathing mode in more detail, we subtract the incoherent background by fitting a single exponential function to the dynamics and subtract this fit from the data. In Fig.~\ref{fig:2DPlot}, we then plot the coherent contribution to the magnetization dynamics in dependence of both pump-pump and pump-probe delay. As already seen in Fig.~\ref{fig:DoublePumpSignal}a) for selected phase delays, the B/SK breathing depends on the pump-pump double-pulse excitation scheme. We find delays at which the breathing mode is present or even enhanced (marked yellow), and delays at which the breathing mode is suppressed (marked black). Interestingly, there is a phase shift of the B/SK breathing mode with respect to the pump-pump delay. 

\begin{figure}[htb]
     \centering
     \includegraphics[width=\columnwidth]{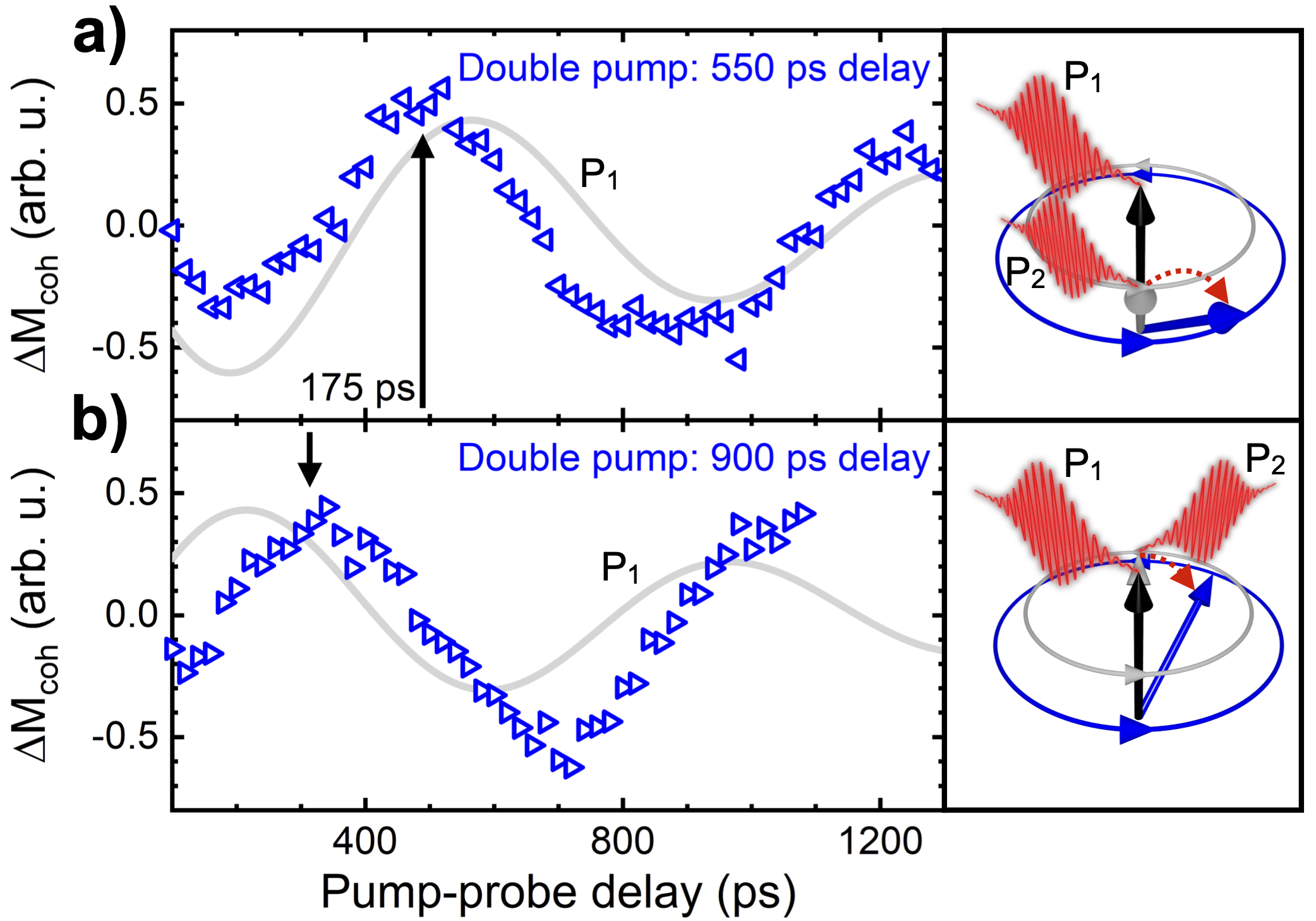}
     \caption{\textbf{Left:} Coherent magnetization change induced by double excitation (blue triangles) considering a pump-pump delay of a) 550\,ps and b) 900\,ps in comparison to a fit of the coherent $P_1$ single excitation response (gray line). The pump-probe delay is referenced to the second excitation $P_2$. In this case, the difference in pump-pump delay accounts for a phase shift of $\Delta t=175$\,ps ($\approx \pi/2$). \textbf{Right:} Phasor representation of the magnetization at the corresponding pump-pump delays indicating a magnetization precession with identical amplitude, but different phase.}
     \label{fig:Phase}
\end{figure}

In Fig.~\ref{fig:Phase}, we analyze this phase shift in more detail and plot two selected double-pump induced breathing modes (blue triangles in a and b) that exhibit a phase shift of approximately $\pi/2$, while the amplitude remains almost unchanged. To explain the origin of the phase shift, we again model the response to the double excitation using the phasor representation of the magnetization vector. Magnetization precession is started by the pump excitation due to the laser-induced anisotropy change (gray arrow, see also Fig.~\ref{fig:DoublePumpSignal}b I). A second excitation induces a similar anisotropy change, i.e., the magnetization is tilted in the same direction as in the case of single excitation (blue arrow tilted right). For this reason, two points in time exist (here: $\approx T/4$ and $\approx 3T/4$) at which the second excitation results in an identical precession amplitude as given by the blue circle, however with a different phase. Thus, an adjustment of fluence and pump-pump delay allows for precise tuning of both phase and amplitude of the B/SK breathing mode.

\begin{figure}[htb]
     \centering
     \includegraphics[width=\columnwidth]{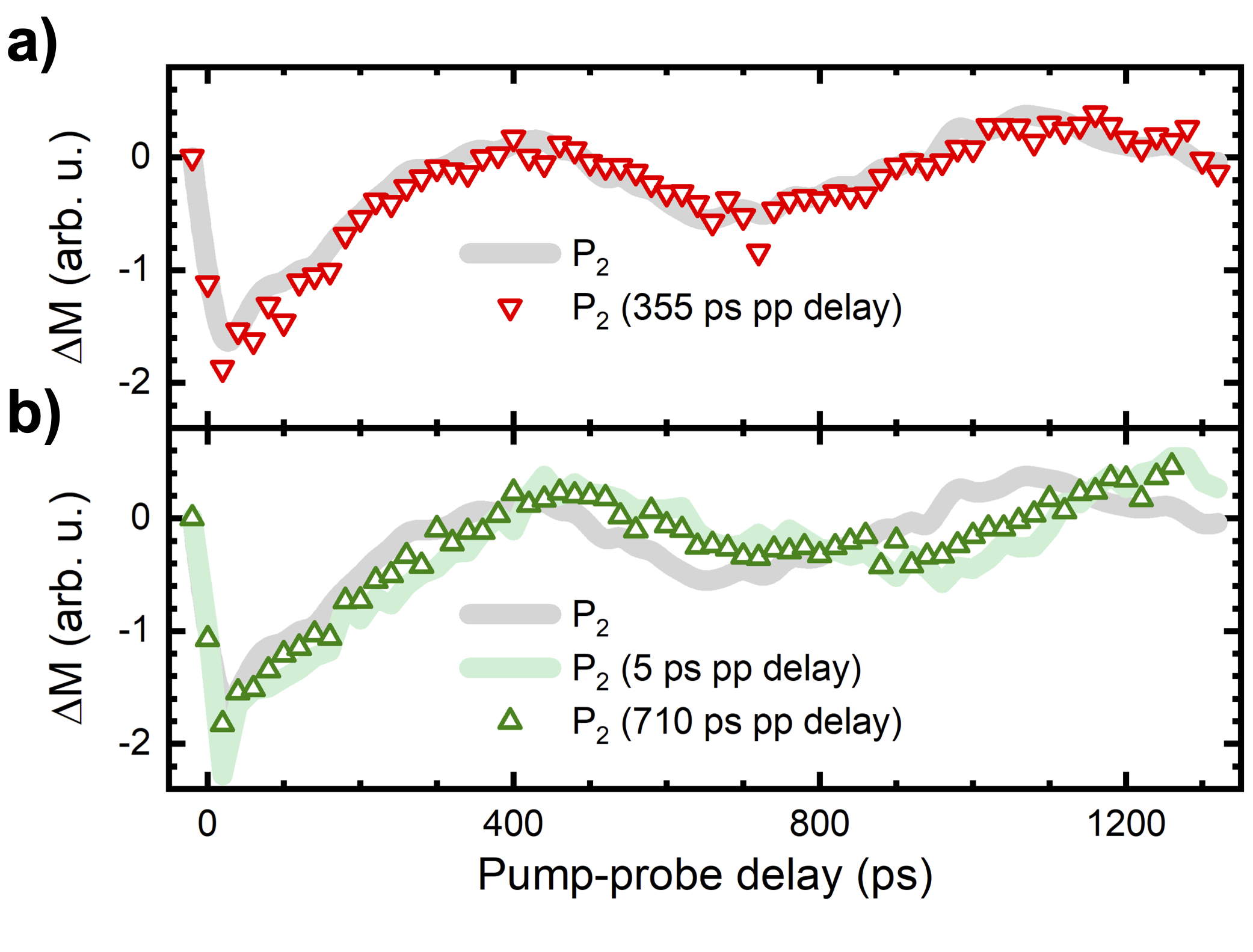}
     \caption{Calculated magnetization dynamics induced by the second pump excitation at different pump-pump delays \textbf{a)} 355\,ps, \textbf{b)} 5\,ps and 710\,ps, respectively. The pump-probe delay is referenced to the second excitation $P_2$. The gray line denotes the response to the plain second excitation $P_2$.}
     \label{fig:Pump2Signal}
\end{figure}

What we have not yet analyzed is to what extent the response of the B/SK lattice to the second excitation $P_2$ depends on the instantaneous phase of the breathing mode that was started by the first excitation pulse $P_1$. In order to study such nonlinear behavior, we subtract the dynamics induced by $P_1$ from the double-pump data set, taking into account the respective pump-pump delay. The result is the dynamics induced by the second excitation only, if we would assume a simple linear superposition of the response of both individual excitations. This data can directly be compared to the response to the plain $P_2$ signal, depicted in Fig.~\ref{fig:Pump2Signal} as a gray line ($P_2$)~\cite{Note1}. 

As can be seen in Fig.~\ref{fig:Pump2Signal}a), a second excitation at 355\,ps pump-pump delay corresponding to a $\pi$ phase shift (red triangles) induces the exact same dynamics as a single excitation (gray line). We note that the second pulse excites at a point in time at which the breathing mode oscillation almost reaches a local maximum, i.e., the size of the spin objects is close to minimal. This in turn shows that the initial state at $t=0$ is given by spin objects of minimal size. 

In contrast, the response of a second excitation at 710\,ps differs from the single excitation, which is interesting considering that the spin objects are of maximum size at 710\,ps. Therefore, the dynamic state of the spin objects strongly influences the dynamics induced by a second excitation. Here, we find very good agreement to the response at a $P_1-P_2$ delay of 5\,ps. We therefore infer that the size of the spin objects is increased by laser excitation, presumably on the time scale of ultrafast demagnetization, which marks the starting point of the breathing mode. Another excitation of the maximum sized spin objects, whether at $t=5$\,ps or $t=710$\,ps delay leads to a further increase of the core diameter. As a result, the breathing mode frequency softens comparable to an excitation of increased strength~\cite{Titze2024}.

In summary, we studied the response of a B/SK lattice to ultrafast double pump excitation. We were able to control the B/SK breathing mode for different time delays between two excitation pulses. We either amplify or attenuate the breathing, depending on whether the spin objects are in a state of maximum or minimum size at the time of the second excitation, which is of high interest considering novel magnonic devices. Furthermore, we achieved control over the phase of the breathing mode by adjusting the pump-pump delay. Here, one can think of utilizing double excitation of B/SKs as phase shifter for spin waves. In addition, tuning of desired frequencies is possible by both fluence adjustment and sample design, e.g. by tuning the size of the spin objects. Careful analysis reveals that the origin of the B/SK breathing mode is an ultrafast increase in size of the spin objects, presumably occurring on a picosecond timescale. These results highlight that optical excitation can be used to precisely tune topologically protected spin states on ultrafast timescales. 

We thus demonstrated that double pump excitation of trivial (bubbles) and topological (skyrmions) spin objects is a powerful tool to control their breathing dynamics in both amplitude and phase, opening up new pathways for spin texture-based applications in spintronics.

\acknowledgements
T.T. and D.S. thank Sabri Koraltan and Henning Ulrichs for helpful discussion. T.S. and M.A. gratefully acknowledge funding from Deutsche Forschungsgemeinschaft (DFG, German Research Foundation) grant no.\,507821284. D.S., S.M.\,and T.T.\,acknowledge funding by the DFG, grant no.\,217133147/SFB 1073, project A02. 

\bibliographystyle{apsrev4-2}
%

\end{document}